\documentclass[11pt]{article}
\pdfoutput=1

\usepackage{amsmath,amsthm,amsfonts,amssymb,srcltx,empheq}
\usepackage{graphicx}
\usepackage[all]{xy}
\usepackage{color}
\usepackage{longtable}
\usepackage{hhline}

\usepackage{caption}
\theoremstyle{plain}

\theoremstyle{remark}

\renewcommand{\thefootnote}{\fnsymbol{footnote}}

\makeatletter
\def\Left#1#2\Right{\begingroup%
   \def\ts@r{\nulldelimiterspace=0pt \mathsurround=0pt}%
   \let\@hat=#1%
   \def\sht@im{#2}%
   \def\@t{{\mathchoice{\def\@fen{\displaystyle}\k@fel}%
          {\def\@fen{\textstyle}\k@fel}%
          {\def\@fen{\scriptstyle}\k@fel}%
          {\def\@fen{\scriptscriptstyle}\k@fel}}}%
   \def\g@rin{\ts@r\left\@hat\vphantom{\sht@im}\right.}%
   \def\k@fel{\setbox0=\hbox{$\@fen\g@rin$}\hbox{%
      $\@fen \kern.3875\wd0 \copy0 \kern-.3875\wd0%
      \llap{\copy0}\kern.3875\wd0$}}%
      \def\pt@h{\mathopen\@t}\pt@h\sht@im%
      \Right}%
\def\Right#1{\let\@hat=#1%
   \def\st@m{\mathclose\@t}%
   \st@m\endgroup}
\makeatother

\makeatletter
 \renewcommand{\theequation}{%
       \thesection.\arabic{equation}}
 \@addtoreset{equation}{section}
\makeatother

\makeatletter
\def\eqnarray{%
 \stepcounter{equation}%
 \let\@currentlabel=\theequation
 \global\@eqnswtrue
 \global\@eqcnt\z@
 \tabskip\@centering
 \let\\=\@eqncr
 $$\halign to \displaywidth\bgroup\@eqnsel\hskip\@centering
 $\displaystyle\tabskip\z@{##}$&\global\@eqcnt\@ne
 \hfil$\displaystyle{{}##{}}$\hfil
 &\global\@eqcnt\tw@$\displaystyle\tabskip\z@{##}$\hfil
 \tabskip\@centering&\llap{##}\tabskip\z@\cr}
\makeatother

\begin{document}
\begin{titlepage}

\hfill{KEK-TH-2160}

\begin{center}
\vspace*{1cm}
{\Large \bf
F-theory Flux Vacua and Attractor Equations
}
\vskip 1.5cm
{\large Yoshinori Honma${}^{a}$\footnote[2]{yhonma@law.meijigakuin.ac.jp} and 
Hajime Otsuka${}^b$\footnote[3]{hotsuka@post.kek.jp}}
\vskip 1.0cm
{\it 
${}^a$%
Department of Current Legal Studies, \\
Meiji Gakuin University, \\
Yokohama, Kanagawa 244-8539, Japan \\
\vspace*{0.5cm}
${}^b$%
Theory Center, Institute of Particle and Nuclear Studies, \\
High Energy Accelerator Research Organization, KEK, \\
1-1 Oho, Tsukuba, Ibaraki 305-0801, Japan\\}
\end{center}
\vskip2.5cm

\begin{abstract}
We examine the vacuum structure of 4D effective theories of moduli fields 
in spacetime compactifications with quantized background fluxes. Imposing the 
no-scale structure for the volume deformations, we numerically investigate the 
distributions of flux vacua of the effective potential in complex structure 
moduli and axio-dilaton directions for two explicit examples in Type IIB string 
theory and F-theory compactifications. It turns out that distributions of 
non-supersymmetric flux vacua exhibit a non-increasing functional behavior of 
several on-shell quantities with respect to the string coupling.
We point out that this phenomena can be deeply connected with a previously-reported 
possible correspondence between the flux vacua in moduli stabilization problem and the 
attractor mechanism in supergravity, and our explicit demonstration implies that 
such a correspondence generically exist even in the framework of F-theory. In 
particular, we confirm that the solutions of the effective potential we explicitly 
evaluated in Type IIB and F-theory flux compactifications indeed satisfy the 
generalized form of the attractor equations simultaneously.
\end{abstract}
\end{titlepage}

\renewcommand{\thefootnote}{\arabic{footnote}} \setcounter{footnote}{0}

\section{Introduction}\label{sec:introduction}

Stabilization of extra massless scalar moduli fields of 4D effective theories arising 
from spacetime compactifications is of particular interest in the field of string theory 
and its broad applications to cosmology and phenomenology. In particular, remarkable 
progress for cosmological observations in recent years has motivated us to realize de 
Sitter spacetime with stabilized moduli in the framework of string theory and contrive a 
concrete setup in which basic quantum gravitational issues can be thoroughly investigated. 
In the study of this moduli stabilization problem, one looks at the scalar potential of 
4D ${\cal{N}}=1$ effective theories arising from spacetime compactifications. In the 
language of 4D ${\cal{N}}=1$ supersymmetry, there are two kinds of contributions to the 
scalar potential of moduli fields, namely the K$\ddot{{\textrm{a}}}$hler potential and 
the superpotential. String compactifications enables us to derive these quantities quantum 
mechanically from the geometry of internal compact spaces, often taken to be a Calabi-Yau 
manifold to use sufficient amount of supersymmetry while keeping the problem nontrivial. 
Inclusion of background fluxes is a crucial ingredient to generate interactions between 
moduli fields in superpotential, leading to their stabilization in a controllable way.

To grasp the background and clarify our motivation for the present paper in more detail, 
let us begin with the 4D effective action arising from Type IIB Calabi-Yau orientifolds 
or the F-theory on Calabi-Yau fourfolds \cite{Vafa:1996xn}, both of which can be described 
by the 4D ${\cal{N}}=1$ supergravity language. Suppose the K$\ddot{{\textrm{a}}}$hler moduli 
fields labeled by $\alpha$ are absent in the superpotential $W$, and also the 
K$\ddot{{\textrm{a}}}$hler potential $K$ satisfies the condition $K^{\alpha \bar{\alpha}} 
\partial_{\alpha} K \partial_{\bar{\alpha}} \overline{K}=3$, the $F$-term scalar potential 
of moduli fields becomes a no-scale type
\begin{align}
V = e^K\left( K^{I \bar{J}} D_I W D_{ \bar{J}} \overline{W} \right),
\label{nosv}
\end{align}
in the reduced Planck mass unit $M_{\rm Pl}=1$. Here $I,J$ run over the complex structure 
moduli of a Calabi-Yau threefold and the axio-dilaton in a Type IIB setup, which can be 
also regarded as a unified complex structure moduli of a Calabi-Yau fourfold from the 
F-theory perspective. Here $D_{I} \equiv \partial_I + (\partial_I K)$ denote the 
K$\ddot{{\textrm{a}}}$hler covariant derivative and $K^{I \bar{J}}$ is the inverse of the 
K$\ddot{{\textrm{a}}}$hler metric $K_{I \bar{J}} = \partial_I \partial_{\bar{J}} K$. 
Throughout this paper, we focus on this no-scale type potential.

It is well known that the imaginary self-dual three-form fluxes in Type IIB string theory 
or the self-dual four-form fluxes in F-theory can generate suitable flux superpotential 
$W$ to stabilize moduli fields at $D_I W=0$, leading to stable Minkowski 
minima \cite{Giddings:2001yu,Gukov:1999ya,Dasgupta:1999ss}. While this prescription completely 
fix the complex structure moduli and axio-dilaton, the K$\ddot{{\textrm{a}}}$hler moduli of 
the compactification remain unfixed. In order to fix all the moduli and construct realistic 
cosmologies describing the accelerated expansion of the universe, quantum $\alpha'$-corrections 
and non-perturbative effects should be further included. This issue has been resolved in the 
explicit constructions of the Kachru-Kallosh-Linde-Trivedi (KKLT) model \cite{Kachru:2003aw} 
and the LARGE Volume Scenario (LVS) \cite{Balasubramanian:2005zx,Conlon:2005ki}. There the 
small value of de Sitter cosmological constant for the late-time cosmology has been realized 
by an inclusion of the anti-D3-brane inducing an uplifting potential.

On the other hand, as has been pointed out for instance in \cite{Saltman:2004sn}, effective 
theories equipped with the no-scale type potential can also accommodate supersymmetry-breaking 
minima with $V \neq 0$ in complex structure moduli and axio-dilaton directions. There the 
spontaneous supersymmetry breaking effect due to a small vacuum energy can be regarded as a 
variant of the anti-D-branes to produce tiny cosmological constant. In the same spirit, 
several supersymmetry-breaking scenarios has been discussed, for instance, 
in \cite{Gallego:2017dvd,Blaback:2015zra}. While such an uplifting mechanism may have broad 
applications to de Sitter model building, comprehensive study has not been fully elucidated, 
possibly due to the complexity of analysis of the non-supersymmetric vacua. One of our goals 
in the present paper is to fill this gap and understand the characteristics of the flux vacua 
in general compactifications.

Specifically, we investigate the vacuum structure of two explicit examples in Type IIB string
theory and F-theory compactifications while assuming the no-scale structure (\ref{nosv}).\footnote{Note that, as we will discuss later, the no-scale analysis which ignores the K\"{a}hler moduli stabilization does not support the accelerated expansion in itself, and further modifications of the models are required if one wants to find out a landscape of explicit de Sitter minima following our approach.} 
By solving the system numerically and finding the non-supersymmetric flux vacua in both setups, 
we find that several on-shell quantities exhibit a non-increasing functional 
behavior with respect to the string coupling. In the process of clarifying an underlying dynamics of 
this phenomena, we find that this characteristic behavior of flux vacua can be regarded as a 
direct consequence of a previously-studied possible correspondence between moduli stabilization 
problem and attractor mechanism in supergravity emphasized in \cite{Kallosh:2005ax} (see 
also \cite{Kallosh:2005bj,Kallosh:2006bt,Alishahiha:2006jd,Bellucci:2007ds,Larsen:2009fw}), and our 
demonstration provides a nontrivial supporting evidence for this kind of correspondence. 
Especially, one can check that our numerical solutions as well as analytic Minkowski solutions 
indeed satisfy the appropriately generalized attractor equation simultaneously, even in the 
framework of F-theory.

This paper is organized as follows. We first take a brief look at an effective theory arising 
from Type IIB toroidal orientifold compactification and study its vacuum structure in Section 
\ref{sec:typeIIB}. In Section \ref{sec:Fth}, we move on to analyze the effective potential of 
moduli fields arising from F-theory compactified on a Calabi-Yau fourfold. Finally in 
Section \ref{sec:Att}, we reconsider our results from another viewpoint and confirm that the 
moduli stabilization problem discussed in Sections \ref{sec:typeIIB} and \ref{sec:Fth} can be 
rephrased in terms of the attractor mechanism in supergravity. Section \ref{sec:conclusion} 
is devoted to conclusions and discussions. In Appendix \ref{app:attF}, we describes the F-theory 
generalization of the attractor equation.

\section{Type IIB Toroidal Orientifold}\label{sec:typeIIB}

First we consider a well-known Type IIB toroidal orientifold model and investigate its vacuum 
structure. Explicit demonstrations in this simple setup would make the arguments in subsequent 
sections more intelligible.

\subsection{Setup}\label{subsec:example1}

Here we briefly look at the effective potential of moduli fields in Type IIB string theory 
compactified on a $T^6/\mathbb{Z}_2$ toroidal orientifold. We refer the reader 
to \cite{Kachru:2002he,Frey:2002hf} for the details about the construction.

Let us represent the periodic six real coordinates on $T^6$ by $x^{i}$ and $y^i$ with 
$i,j=1,2,3$, and take the holomorphic one-forms of the $T^2$ submanifolds as 
$dz^i=dx^i +\tau^{ij}dy^j$. By choosing the orientation as $\int_{T^6} dx^1\wedge dx^2\wedge 
dx^3 \wedge dy^1 \wedge dy^2 \wedge dy^3=1$ and representing the three-form cohomology basis 
in $H^3(T^6, \mathbb{Z})$ as
\begin{align}
\begin{split}
  \alpha_0 &= dx^1 \wedge dx^2 \wedge dx^3, \ \ \ \ \ \ \ \ \ \ \ \alpha_{ij} = \frac{1}{2} 
  \epsilon_{ikl} dx^k \wedge dx^l \wedge dy^j, \\ \beta^{ij} &= -\frac{1}{2} \epsilon_{jkl} 
  dy^k \wedge dy^l \wedge dx^i, \ \ \ \ \beta_0 = dy^1 \wedge dy^2 \wedge dy^3,
\end{split}
\end{align}
the holomorphic three-form $\Omega \equiv dz^1 \wedge dz^2 \wedge dz^3$ on $T^6/\mathbb{Z}_2$ 
can be described as
\begin{align}
\Omega =\alpha_0 +\alpha_{ij}\tau^{ij} -\frac{1}{2}\beta^{ij}(\epsilon_{ikl}
\epsilon_{jmn}\tau^{km}\tau^{ln})+\beta^0 ({\rm det}\tau).
\end{align}
Here the matrices $\tau^{ij}$ describe the complex structure moduli of the geometry and the 
three-form cohomology basis is normalized as $\int_{T^6} \alpha_I \wedge \beta^J =\delta_I^J$. 
The three-form fluxes $F_3$ and $H_3$ in RR and NSNS sectors in Type IIB string theory can be 
expanded in the same basis as
\begin{align}
\begin{split}
&\frac{1}{(2\pi)^2 \alpha'}F_3 = a^0 \alpha_0 +a^{ij}\alpha_{ij} +b_{ij}\beta^{ij} +b_0 \beta^0,
\\
&\frac{1}{(2\pi)^2 \alpha'}H_3 = c^0 \alpha_0 +c^{ij}\alpha_{ij} +d_{ij}\beta^{ij} +d_0 \beta^0,
\end{split}
\end{align}
where the coefficients $a^0,a^{ij},b_0,b_{ij},c^0,c^{ij},d_0,d_{ij} \in 2\mathbb{Z}$ 
describe the quanta of background fluxes. Note that here we focus on even numbers of flux 
quanta to prohibit the appearance of delicate exotic O3-planes \cite{Frey:2002hf}.

The explicit form of the K$\ddot{{\textrm{a}}}$hler potential for moduli fields is given by 
\begin{align}
K=-\ln{\left[ -i(S-\overline{S}) \right]} -\ln{\left[ -i\int \Omega \wedge \bar{\Omega} 
\right]} -2\ln{\cal{V}},
\label{torK}
\end{align}
where $S$ denotes the axio-dilaton. The hermitian norm of $\Omega$ corresponds to the 
Weil-Petersson metric of the complex structure moduli space and ${\cal V}$ is a volume of 
the internal manifold $T^6/\mathbb{Z}_2$ in the Einstein-frame, measured in units of 
$2\pi \sqrt{\alpha^\prime}$. Plugging (\ref{torK}) into the scalar potential (\ref{nosv}), 
one finds that $V$ no longer depends on the K$\ddot{{\textrm{a}}}$hler moduli fields 
except for a prefactor $\frac{1}{{\cal V}^2}>0$ treated as a constant throughout this paper.
On the other hand, the superpotential induced by background fluxes in Type IIB 
compactifications\footnote{Here we use a convention where the factor $(4\pi)^{-1/2}$ appear 
in the superpotential to match with the dimensional reduction of Type IIB supergravity 
action.} takes the following form \cite{Gukov:1999ya}
\begin{align}
W = \frac{1}{\sqrt{4\pi}(2\pi)^2 \alpha'}\int \Omega \wedge G_3,
\end{align}
where $G_3 \equiv F_3-SH_3$ denotes the combined three-form fluxes. On $T^6/\mathbb{Z}_2$ 
geometry, this becomes
\begin{align}
\begin{split}
W=\frac{1}{\sqrt{4\pi}} \Bigg( &(a^0 -Sc^0){\rm det}\tau -\frac{1}{2}(a^{ij} -Sc^{ij})
\epsilon_{ikm}\epsilon_{jpq}\tau^{kp}\tau^{mq} \\
& -(b_{ij} -Sd_{ij})\tau^{ij} -(b_0 -Sd_0) \Bigg) .
\end{split}
\end{align}
Recalling that there exist $2^6$ O3-planes on $T^6/{\mathbb Z}_2$ orientifold, the tadpole 
cancellation condition
\begin{align}
32 -2n_{{\rm D}3}= \frac{1}{(2\pi)^4 (\alpha')^2}\int H_3 \wedge F_3 = c^0b_0 +c^{ii}b_{ii} 
-d_{ii}a^{ii} -d_0a^0,
\label{tadcan}
\end{align}
must be also satisfied to ensure the global conservation of fluxes inside the compact manifold.
Here $n_{{\rm D}3}$ denotes the number of mobile D3-branes, which is set to be zero in the 
following discussions.

To simplify the analysis, we further focus on an isotropic case for the system, namely we only 
consider diagonal components of $\tau^{ij}$ and fluxes satisfying the following condition
\begin{align}
\begin{split}
\tau^{11} &=\tau^{22} =\tau^{33} \equiv \tau, \\
a^{11} &=a^{22}=a^{33} \equiv a^1, \\
b_{11} &=b_{22}=b_{33} \equiv b_1, \\
c^{11} &=c^{22}=c^{33} \equiv c^1, \\
d_{11} &=d_{22}=d_{33} \equiv d_1,
\end{split}
\end{align}
as well as the fluxes $\{ a^0, b_0, c^0, d_0 \}$. Then the K$\ddot{{\textrm{a}}}$hler potential 
(\ref{torK}) of the $T^6/{\mathbb Z}_2$ orientifold model reduces to be
\begin{align}
K=-\ln{\left[ (S-\overline{S})(\tau-\overline{\tau})^3 \right]}-2\ln{\cal{V}},
\label{isok}
\end{align}
and the flux-induced superpotential becomes
\begin{align}
W = (4\pi)^{-1/2}\left( (a^0 -S c^0) \tau^3 -3 (a^1 -S c^1) \tau^2 -3(b_1 -S d_1)\tau 
-(b_0 - S d_0)\right).
\label{isow}
\end{align}

Combining the above setup with the definition of the no-scale type potential (\ref{nosv}), 
one can study the vacuum structure of the model explicitly. Here we consider a condition
\begin{align}
a^0=c^1=b_1=d_0=0,
\end{align}
and non-zero otherwise, which corresponds to the imaginary self-dual condition allowing (2,1) 
and (0,3) piece for $F_3$ and the imaginary anti-self-dual condition allowing (3,0) and (1,2) 
piece for $H_3$. This asymmetric choice of background fluxes was chosen to completely fix the 
complex structure moduli and axio-dilaton, coming from the fact that the $T^6/{\mathbb Z}_2$ 
toroidal orientifold model originally preserves 4D ${\cal{N}}=4$ supersymmetry.\footnote{See for 
instance \cite{DAuria:2002qje} about 4D ${\cal{N}}=4$ gauged supergravity description of the 
model.} In this setup, the model has a Minkowski minimum\footnote{In order to be supersymmetric, 
the minimum also need to satisfy $W=0$ and allowed fluxes are further constrained.} as a solution 
to the $F$-term conditions $D_I W=0$, where $V$ becomes zero and the values of moduli fields are 
fixed as
\begin{align}
{\textrm{Re}}\tau ={\textrm{Re}}S=0, \ \ \ \ {\textrm{Im}}\tau =\left(\frac{b_0 {d_1}'}{a^1 c^0}
\right)^{1/4}, \ {\textrm{Im}} S =\left( \frac{(a^1)^3 b_0}{c^0 {d_1}'^3}\right)^{1/4},
\label{torvacuum}
\end{align}
where ${d_1}' \equiv -d_1$. On the other hand, when we consider a condition
\begin{align}
c^0=a^1=d_1=b_0=0,
\end{align}
and non-zero otherwise, this choice corresponds to the imaginary self-dual condition for $H_3$ and 
the imaginary anti-self-dual condition for $F_3$. Then the model has a Minkowski minimum at which 
the values of moduli fields are fixed as
\begin{align}
{\textrm{Re}}\tau &={\textrm{Re}}S=0, \ \ \ \ {\textrm{Im}}\tau =\left(\frac{{b_1}' {d_0}'}{a^0 
{c^1}'}\right)^{1/4}, \ {\textrm{Im}}S=\left( \frac{a^0 ({b_1}')^3}{({c^1}')^3 {d_0}'}\right)^{1/4},
\label{torvac2}
\end{align}
where ${b_1}'\equiv -b_1, {c^1}'\equiv -c^1$ and ${d_0}' \equiv -d_0$. Later we will utilize these 
simple analytic solutions (\ref{torvacuum}) and (\ref{torvac2}) when we discuss about a 
correspondence between the moduli stabilization and a seemingly different topic in supergravity.

\subsection{Distribution of non-supersymmetric flux vacua}
\label{subsec:resultIIB}

Having introduced the setup of the $T^6/\mathbb{Z}_2$ toroidal orientifold model, now we turn to 
analyze the non-supersymmetric flux vacua with fixed moduli $\Phi_I \equiv \{ \tau^{ij}, S \}$, 
while taking into account the tadpole cancellation condition for quantized background fluxes 
appropriately. 

As emphasized in \cite{Saltman:2004sn}, the no-scale type potential (\ref{nosv}) can also 
accommodate the supersymmetry-breaking minima characterized by
\begin{align}
(\partial_I V)|_{\Phi_I^0} =0, \qquad
V|_{\Phi_I^0} =\frac{\epsilon}{{\cal V}^2}>0,
\label{nonsv}
\end{align}
with the mass squared of the moduli fields being strictly positive. Note that the nonzero 
vacuum energy at $\Phi_I^0$ is proportional to a constant Calabi-Yau volume arising from 
the prefactor $e^K$ in the scalar potential, as mentioned in the previous section. Numerical 
constant $\epsilon$ determined by fluxes can take sufficiently small value owing to the 
richness of the choice of background fluxes, and this becomes a candidate of the source of 
an uplifting potential. Throughout this paper, we only pick up the solutions satisfying
$\epsilon <1$ as true flux vacua, otherwise the effective theory description of the system 
turns out to be unreliable.

In our numerical analysis, we utilize the ``FindMinimum'' function in Mathematica to find 
supersymmetry-breaking minima of the effective potential.\footnote{More precisely, we 
performed the analysis with working precision of 50 digits and accuracy goal of 20 digits. 
We also checked that our results are all insensitive to such parameter choices and pick up 
the accurate solutions only.} To simplify the analysis, let us again focus on the isotropic 
case for $\tau^{ij}$ and background fluxes in which the the K$\ddot{{\textrm{a}}}$hler 
potential and the superpotential are given by (\ref{isok}) and (\ref{isow}), respectively.
Within a flux range $-12\leq a^0,a^1,b_0,b_1,c^0,c^1,d_0,d_1\leq 12$ while imposing the 
tadpole cancellation condition (\ref{tadcan}) with $n_{\rm D3}=0$, we found 962 
supersymmetry-breaking minima with no tachyonic instability. 

\begin{figure}[ht]
\begin{minipage}{0.5\hsize}
\begin{center}
\includegraphics[width=80mm]{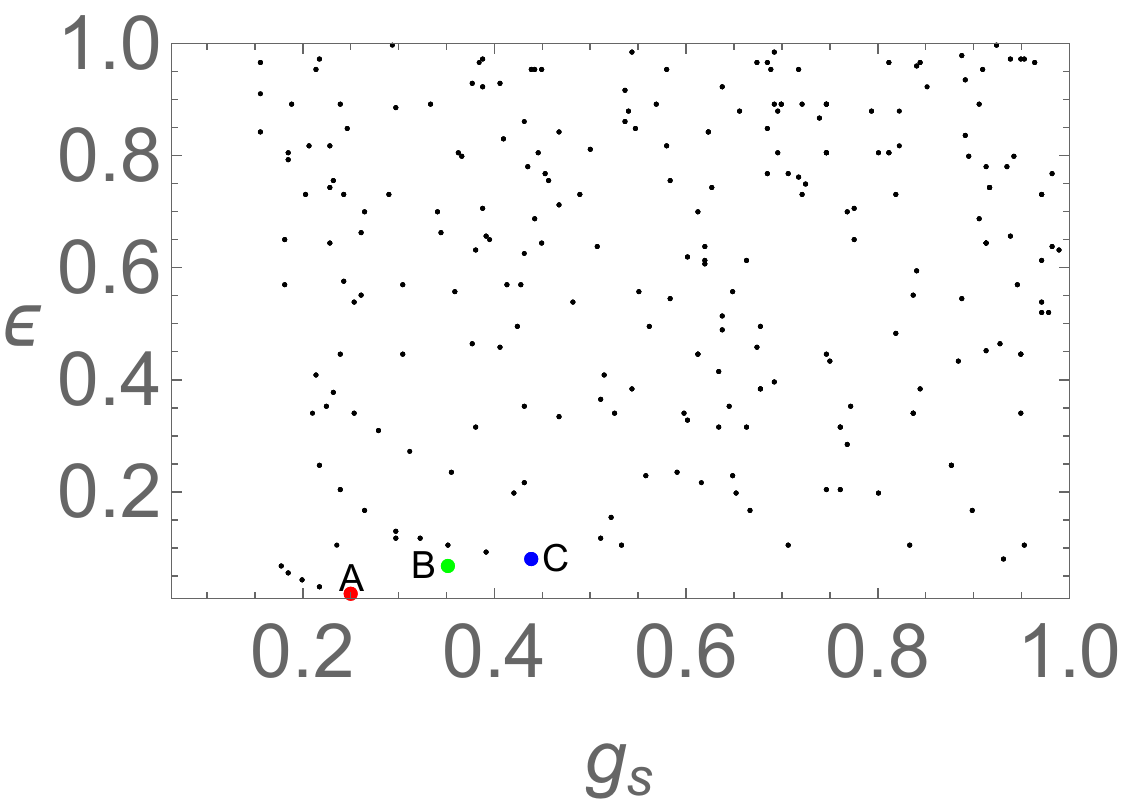}
\end{center}
\end{minipage}
\begin{minipage}{0.5\hsize}
\begin{center}
\includegraphics[width=80mm]{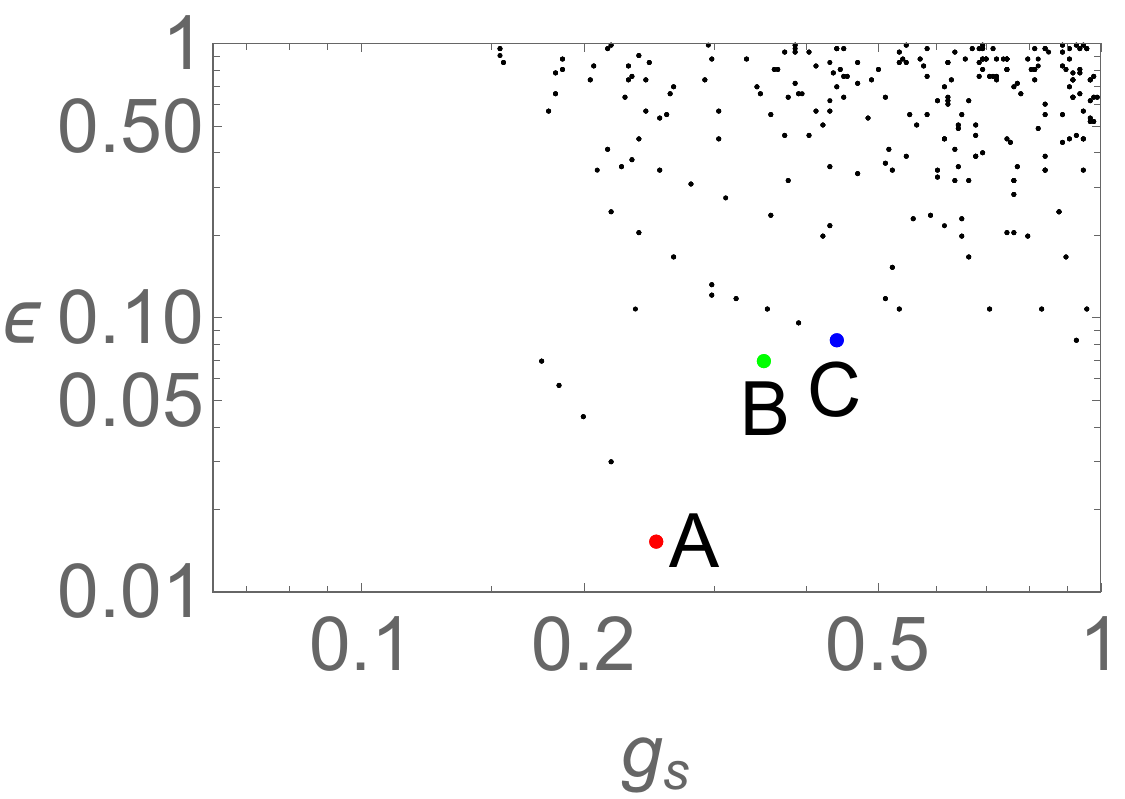}
\end{center}
\end{minipage}
\caption{Numerical results for the distribution of flux vacua with nonzero vacuum energy. 
Each dot corresponds to a solution with different set of fluxes. Note that, however, 
generically these dots can be degenerated in this 2D plot, due to the existence of another 
direction of the stabilized moduli $\tau$ which is not depicted here for the simplicity.
The right figure represents the log-log plot of the same set of solutions.}
\label{fig:T6}
\end{figure}

\begin{figure}[htb]
  \begin{minipage}{0.5\hsize}
   \begin{center}
     \includegraphics[width=80mm]{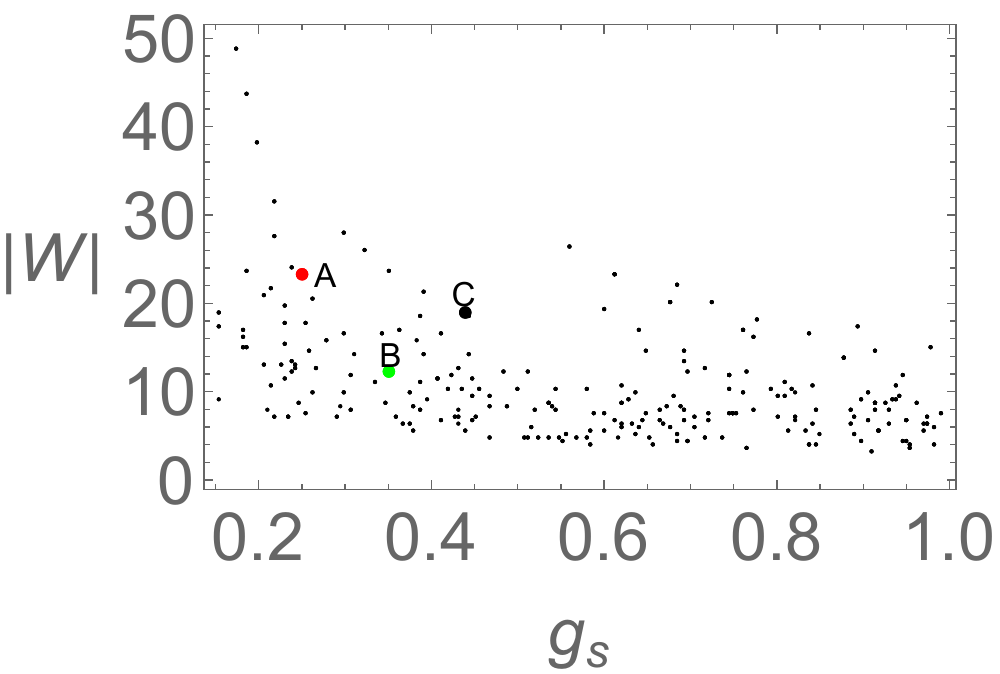}
    \end{center}
   \end{minipage}
  \begin{minipage}{0.5\hsize}
   \begin{center}
    \includegraphics[width=80mm]{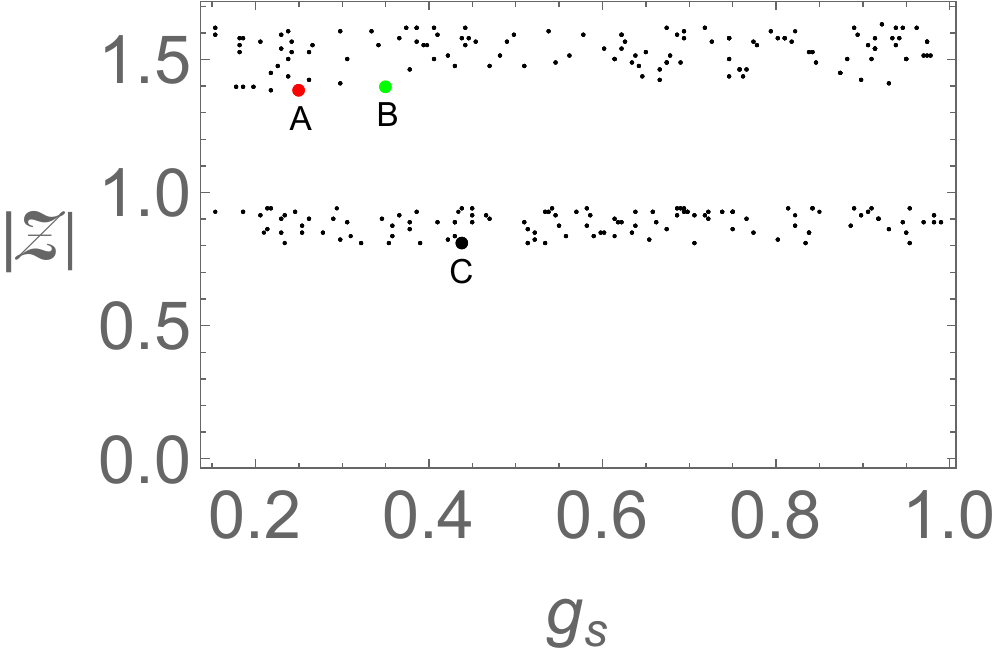}
   \end{center}
  \end{minipage}
    \caption{Numerical results for the absolute values of flux superpotential $|W|$ 
    and a quantity $|{\cal Z}|=e^{K/2}|W|$ at the each of the solutions depicted in 
    Figure \ref{fig:T6}.}
    \label{fig:T3}
\end{figure}

The result is depicted in 
Figure \ref{fig:T6}, where we plot the height of the normalized potential 
$\epsilon ={\cal V}^2V$ with respect to the string coupling $g_s \equiv ({\rm Im}S)^{-1}$ 
at the each of the minima. The other on-shell quantities are also shown in Figure \ref{fig:T3}. 
Note that generically an ambiguity originating from the 
Type IIB SL(2,$\mathbb{Z}$) duality transformation generates physically-equivalent solutions 
and a careful analysis is required. Here and in what follows, we deduct this 
ambiguity\footnote{For the present example, there exists another type of redundancy arising 
from SL(2,$\mathbb{Z}$) $\subset$ SL(6,$\mathbb{Z}$) transformation acting on $T^6$ and we 
have also deducted this ambiguity by requiring $|\textrm{Re} \tau| \leq 1/2$ and $|\tau| 
\geq 1$.} by selecting only the solutions satisfying $|\textrm{Re} S| \leq 1/2$ and $|S| 
\geq 1$ while keeping the string coupling small but finite: $0 < g_s <1$.

\clearpage
To provide typical examples of our numerical results, here we pick up three independent 
flux vacua determined under the following choice of fluxes:
\begin{table}[ht]
    \centering
    \begin{tabular}{|c|c|} \hline
   Vacuum    & Set of fluxes $(a^0, a^{1},b_0, b_1, c^0, c^1, d_0, d_1)$ \\ \hline \hline
   A   & $(-2, 4, 0, -4, 0, 0, 4, -2)$ \\ \hline
   B    &  $(4, -4, 2, 2, 0, 0, -2, 2)$ \\ \hline
   C    & $(-2, 2, 4, 4, 0, 0, 10, -2)$ \\  \hline
    \end{tabular}
\end{table}

\noindent The vacuum A, B and C are represented in Figure \ref{fig:T6} by the red, green and blue 
points, respectively. Explicit values of stabilized moduli and various on-shell quantities at the 
each of these flux vacua are summarized in Tables~\ref{tab:T61} and~\ref{tab:T62}.

\begin{table}[ht]
    \centering
    \begin{tabular}{|c|c|c|c|c|c|} \hline
   Vacuum     &  $\tau$  & $S$ & $g_s$ & $\epsilon$ & $W$ \\ \hline \hline
   A     & $0.355+1.64 i$ & $0.271 + 3.98 i$ & 0.251 & $1.52\times 10^{-2}$ & $22.6 +5.08i$ \\ \hline
   B     & $0.0468 +1.19 i$ & $0.493 + 2.84 i$ & 0.352 & $6.99\times 10^{-2}$ & $-11.6 -3.90i$ \\ \hline
   C     & $-0.393+2.48 i$ & $-0.440 + 2.29 i$ & 0.438 & $8.23\times 10^{-2}$ & $14.3 +12.6i$ \\ \hline
    \end{tabular}
    \caption{Explicit values of stabilized moduli fields and on-shell quantities in $M_{\rm Pl}=1$ 
    unit.}
    \label{tab:T61}
\end{table}
\begin{table}[ht]
    \centering
    \begin{tabular}{|c|c|} \hline
   Vacuum    & Eigenvalues of mass matrix $\partial_I \partial_J V \times {\cal V}^{2}$
   \\ \hline \hline
   A   &  (3.41, 0.603, 0.222, $1.91\times 10^{-2}$) \\ \hline
   B    &  (6.66, 1.30, 0.324, $6.85\times 10^{-2}$)\\ \hline
   C    & (2.18, 0.690, $7.29\times 10^{-2}$, $9.76\times 10^{-4}$)\\ \hline
    \end{tabular}
    \caption{Mass eigenvalues of normalized scalar potential $\epsilon$ in $M_{\rm Pl}=1$ unit. 
    Note that we have diagonalized the mass matrix ${\cal V}^{2} \partial_I \partial_J V$ by taking 
    the linear combinations of the basis and the eigenvalues are displayed in the descending order.}
    \label{tab:T62}
\end{table}

From Figures \ref{fig:T6} and \ref{fig:T3}, one finds that the flux vacua of the model seems to 
be determined such that several on-shell quantities are governed by a nontrivial dynamics, rather 
than just providing a completely random distribution. Especially, the distribution of a normalized 
superpotential $|{\cal Z}|=e^{K/2}|W|$ indicates a monotonic,  non-increasing functional dependence 
with respect to the string coupling. As we will see in the next section, another concrete example 
based on F-theory flux compactification exhibits this kind of dependence of on-shell quantities 
more clearly. Therefore let us save the arguments for a physical background of this suggestive 
behavior and its consequences for later discussions in the final section, after looking at the 
F-theory example and its vacuum structure.

Here we provide some comments on a further consistency check condition to be satisfied, 
which can be straightforwardly applied to the F-theory compactification described in the 
next section. To ensure the validity of the effective theory description of the system in 
the framework of spacetime compactifications, the vacuum energy at the each of the minima 
needs to be smaller than the mass scales of the Kaluza-Klein and the stringy modes as
\begin{align}
\langle V \rangle \ll m_s^4, m_{\rm KK}^4.
\end{align}
In the Type IIB language, these scales are given by 
$m_s=(\alpha^\prime)^{-1/2} \simeq \frac{g_s^{1/4} \sqrt{\pi}}{ {\cal V}^{1/2}}$ and 
$m_{\rm KK} \simeq \frac{\sqrt{\pi}}{ {\cal V}^{2/3}}$ in the reduced Planck mass unit. 
Therefore the above conditions can be rephrased as a requirement that the dimensionless 
constant $\epsilon$ must satisfy the condition 
\begin{align}
\epsilon \ll {\rm min}({\cal V}^{-2/3}\pi^2 , g_s \pi^2).
\label{eq:epsilon}
\end{align}
For instance if we take a sufficiently large volume ${\cal V}$, e.g. larger than $\simeq 
{\cal{O}}(10^2)$, one can estimate that all the numerical solutions summarized in 
Figure \ref{fig:T6} satisfy (\ref{eq:epsilon}). Of course in order to conduct a detailed 
analysis, ${\cal V}$ needs to be completely fixed by considering an extension of the 
K$\ddot{{\textrm{a}}}$hler moduli sector beyond our no-scale setup\footnote{No-scale 
potential in itself generically provides a runaway potential for the 
K$\ddot{{\textrm{a}}}$hler moduli and does not support the accelerated expansion
\cite{Frey:2002qc,Dine:1985he}, which requires the inclusion of quantum corrections to 
the K$\ddot{{\textrm{a}}}$hler moduli sector. This means that after doing this kind of 
extension in our Type IIB and F-theory setups, the exact relationship between our analysis 
and the de Sitter swampland conjecture \cite{Obied:2018sgi,Garg:2018reu,Ooguri:2018wrx} would come into 
view directly.} in (\ref{torK}).

\section{F-theory Model}
\label{sec:Fth}

Let us move on to discuss the 4D ${\cal N}=1$ effective theory based on F-theory flux 
compactification on a Calabi-Yau fourfold $X_4$. After introducing the setup of a model 
constructed in \cite{Honma:2017uzn}, we numerically analyze the distribution of 
supersymmetry-breaking minima of its scalar potential in the same manner as performed in the 
previous section.

\subsection{Setup}
\label{subsec:setupF}

From the perspective of the F-theory framework, the K$\ddot{{\textrm{a}}}$hler potential for 
complex structure moduli space of a Calabi-Yau fourfold $X_4$ is defined by
\begin{align}
K=-\ln{\int_{X_4}} \Omega \wedge \overline{\Omega},
\label{Kahp}
\end{align}
where $\Omega$ denotes a holomorphic $(4,0)$-form on $X_4$. The F-theory compactification 
also generically admits a superpotential of the form \cite{Gukov:1999ya}
\begin{align}
W = (4\pi)^{-1/2}\int_{X_4} G_4 \wedge \Omega,
\end{align} 
in the presence of four-form fluxes $G_4$, which is inherited from a duality between F-theory 
and M-theory on the same manifold \cite{Becker:1996gj,Sethi:1996es,Haack:2001jz,Denef:2008wq}.
As is the case in the Type IIB orientifold model, background fluxes are required to satisfy 
the tadpole cancellation condition given by 
\begin{align}
\frac{\chi}{24}=n_{\rm{D3}}+\frac{1}{2}\int_{X_4} G_4 \wedge G_4,
\end{align}
in order to be conserved within a compact manifold $X_4$ globally. Here $\chi$ is the Euler 
characteristic of $X_4$ and $n_{\rm{D3}}$ denotes the total number of the mobile D3-branes. 

For a Calabi-Yau fourfold $X_4$ with $h^{3,1}(X_4)$ complex structure moduli, the period 
integrals of holomorphic $(4,0)$-form defined by
\begin{align}
\Pi_i = \int_{\gamma^i} \Omega,
\label{4peri}
\end{align}
encode the moduli dependence of the system, and the K$\ddot{{\textrm{a}}}$hler potential for 
complex structure moduli (\ref{Kahp}) can be rewritten as
\begin{align}
K=-\ln{\left[ \sum_{i,j} \Pi_i \eta^{ij} \overline{\Pi}_j \right]},
\label{Kper}
\end{align}
where $\gamma^i$ with $i=1, \ldots , h^4_H (X_4)$ denote a basis of primary horizontal 
subspace of $H_4 (X_4)$. Here we introduced an intersection matrix $\eta^{ij}$ and a dual 
basis $\hat{\gamma}^i$ in $H^4_H (X_4)$ defined by
\begin{align}
\eta^{ij} = \int_{X_4} \hat{\gamma}^i \wedge \hat{\gamma}^j,
 \ \ \ \ \ \ \ \int_{\gamma^i} \hat{\gamma}^j = \delta^{ij}. 
\label{intm} 
\end{align}
Similarly, when we turn on background $G_4$ fluxes whose integer quantum numbers are given 
by
\begin{align}
n_i = \int_{\gamma^i} G_4,
\label{Gflux}
\end{align}
they generate a superpotential of the following form
\begin{align}
W = (4\pi)^{-1/2} \sum_{i,j} n_i \Pi_j \eta^{ij}.
\label{fluq}
\end{align}

As an explicit example of F-theory compactification, here we consider the background studied 
in \cite{Honma:2017uzn} 
(see also \cite{Alim:2009bx,Grimm:2009ef,Jockers:2009ti,Grimm:2010gk,Honma:2015iza}), 
whose period integrals and topological intersection matrix are given by
\begin{align}
\begin{split}
\Pi_1&=1, \ \Pi_2=z, \ \Pi_3=-z_1, \ \Pi_4=S, \\
\Pi_5&=5Sz, \ \Pi_6=\frac{5}{2}z^2, \ \Pi_7=2z_1^2, \ \Pi_8=-\frac{5}{2}Sz^2-\frac{5}{3}z^3, \\
\Pi_9&=-\frac{2}{3}z_1^3, \ \Pi_{10}=-\frac{5}{6}z^3, \ \Pi_{11}=\frac{5}{6}Sz^3+\frac{5}{12}z^4
-\frac{1}{6}z_1^4,
\end{split}
\label{per}
\end{align}
\begin{align}
\eta=\begin{pmatrix}
0& 0 & 0 & 0 & 1 \\
0 & 0 & 0 & I_3 & 0 \\
0 & 0 & \widetilde{\eta} & 0 & 0 \\
0 & I_3 & 0 & 0 & 0 \\
1 & 0 & 0 & 0 & 0
\end{pmatrix}, \ \ \ \ \ \ 
\widetilde{\eta}=\begin{pmatrix}
0& \frac{1}{5} & 0 \\[5pt]
\frac{1}{5} & \frac{2}{5} & 0 \\[5pt]
0 & 0 & -\frac{1}{4}
\end{pmatrix},
\label{imat}
\end{align}
with the Euler characteristic  $\chi = 1860$. Here the complex structure moduli of the 
fourfold $z,z-z_1,S$ are originated from a bulk quintic modulus, a brane modulus and the 
axio-dilaton in Type IIB description, respectively. Note that here we have picked up the 
leading interactions only. While this choice is sufficient for the current purpose, further 
quantum corrections can be easily calculated by using mirror symmetry technique or 
supersymmetric localization approach as in \cite{Honma:2013hma}.

Substituting the topological data (\ref{per}) and (\ref{imat}) into the generic formula 
(\ref{Kper}), one can obtain the explicit form of the K$\ddot{{\textrm{a}}}$hler potential 
as
\begin{align}
K=-\ln{\left[ -i(S-\overline{S}) \right]} -\ln{\widetilde{Y}}-2\ln{\cal{V}},
\label{FK}
\end{align}
where
\begin{align}
\widetilde{Y}=\frac{5i}{6}(z-\bar{z})^3+\frac{i}{S-\overline{S}}
\left( \frac{5}{12}(z-\bar{z})^4 -\frac{1}{6}(z_1-\bar{z}_1)^4 \right).
\label{FY}
\end{align}
Here we also added the contribution from the classical K$\ddot{{\textrm{a}}}$hler moduli 
sector. The characteristic property of this kind of F-theory construction based on a particular 
class of Calabi-Yau fourfold is that the next-to-leading-order correction with respect to the 
string coupling $g_s = ({\rm Im}S)^{-1}$ has been incorporated, owing to the elliptically fibered 
structure of the background geometry.\footnote{See \cite{Grimm:2009ef,Jockers:2009ti} for a detailed 
analysis and general construction based on mirror symmetry technique with D-branes.} Similarly, 
the flux-induced superpotential is given by
\begin{align}
\begin{split}
W = \frac{1}{\sqrt{4\pi}} & \Bigg( n_{11}+n_{10}S+n_8 z+n_6 S z+\frac{1}{2}\left( n_5+2 n_6 \right) 
z^2 -\frac{5n_4}{6} z^3 -n_2 \left( \frac{5}{2}Sz^2+\frac{5}{3}z^3 \right)  \\
& \ \ -n_9 z_1-\frac{n_7}{2}z_1^2 -\frac{2n_3}{3}z_1^3+n_1 \left( \frac{5}{6}Sz^3+\frac{5}{12}z^4
-\frac{1}{6}z_1^4 \right) \Bigg),
\label{W}
\end{split}
\end{align}
Note that the tadpole cancellation condition needs to be satisfied by background fluxes (\ref{Gflux}) 
as
\begin{align}
\frac{1860}{24} =& \ n_{\rm{D3}}+n_1 n_{11}+n_2 n_8+n_3 n_9+n_4 n_{10}+\left( \frac{n_5+n_6}{5}\right) 
n_6-\frac{n_7^2}{8},
\label{tadcanF}
\end{align}
which requires that $n_7$ must be $2+4k$ with $k \in {\mathbb{Z}}$ and $n_5+n_6$ or $n_6$ must be 
$5k'$ with $k' \in {\mathbb{Z}}$, to preserve the integrality of each of the flux quanta. 

Here let us impose a condition
\begin{align}
n_2=n_3=n_4=n_8=n_9=n_{10}=0,
\end{align}
and non-zero otherwise, which corresponds to consider the self-dual $G_4$ fluxes only. 
In this setup, the model has a Minkowski minimum with $V=0$ as a solution to the $F$-term condition 
and the values of the moduli fields are fixed as \cite{Honma:2017uzn}
\begin{align}
\begin{split}
{\textrm{Re}}z &={\textrm{Re}}z_1={\textrm{Re}}S=0, \\
{\textrm{Im}}z&=\left(\frac{6n_{11}}{5n_1}\right)^{1/4}\frac{2\sqrt{n_6}}
{(8n_6(n_5+n_6)-5n_7^2)^{1/4}},\\ 
{\textrm{Im}}z_1&=\left(\frac{30n_{11}}{n_1}\right)^{1/4}\frac{\sqrt{n_7}}
{(8n_6(n_5+n_6)-5n_7^2)^{1/4}}, \\
{\textrm{Im}}S &=\left(\frac{6n_{11}}{5n_1}\right)^{1/4}\frac{n_5}{\sqrt{n_6}
(8n_6(n_5+n_6)-5n_7^2)^{1/4}}.
\label{vacuum}
\end{split}
\end{align}
Again, later we will utilize this analytic solution to discuss about a correspondence between 
the moduli stabilization problem and a seemingly different topic in supergravity.

\subsection{Distribution of non-supersymmetric flux vacua}
\label{subsec:resultF}

Based on the above setup, we are now ready to investigate the non-supersymmetric flux vacua 
with fixed  moduli $\Phi_I \equiv \{ z, z_1, S \}$ in the framework of F-theory compactification, 
in a similar way as we demonstrated in Section \ref{subsec:resultIIB}. 

In order to find general supersymmetry-breaking minima (\ref{nonsv}) for the no-scale potential
with (\ref{FK}) and (\ref{W}), we again utilize the ``FindMinimum'' function in Mathematica and 
also check the absence of tachyons at thus obtained flux vacua by confirming the mass squared of 
the moduli fields being strictly positive. In the following analysis, we restrict ourselves to 
the case with $n_{\rm D3}=0$ for simplicity, and we only pick up the solutions satisfying 
$\epsilon <1$ as true flux vacua in order to ensure the effective theory description of the system. 
We also need to be careful about the validity of the leading term approximation of the underlying 
fourfold periods (\ref{per}) from which the scalar potential is determined. During our numerical analysis, thereby we further impose the conditions ${\rm Im}\Phi_I > 1$ to ensure that the leading terms of the periods are sufficiently large compared to the possible quantum loop and instanton corrections, and thus all the obtained minima become 
well-defined.\footnote{We have also checked that our numerical results are consistent with large 
complex structure expansion of the periods, whose radius of convergence can be extracted from 
discriminants of the underlying fourfold.}

\begin{figure}[t]
  \begin{minipage}{0.5\hsize}
   \begin{center}
     \includegraphics[width=80mm]{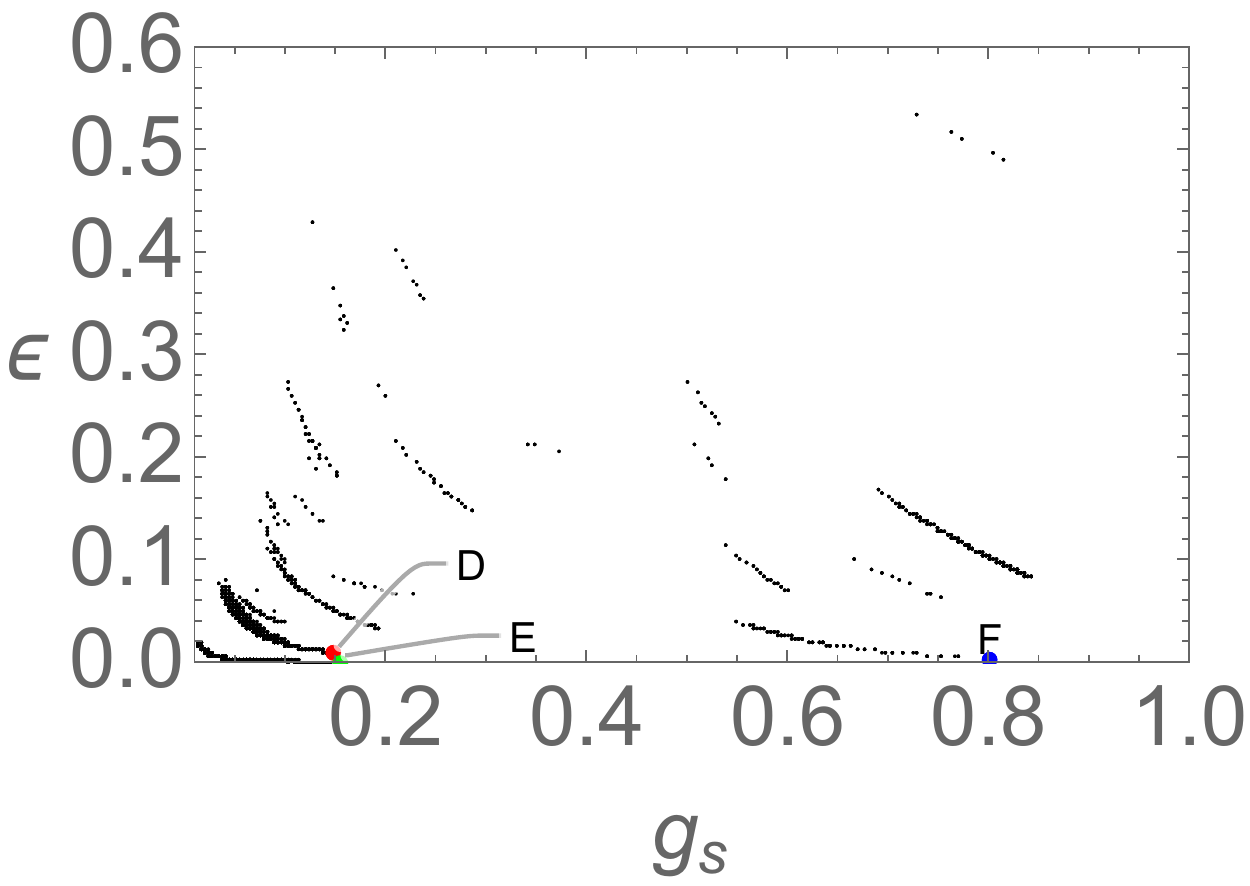}
    \end{center}
   \end{minipage}
  \begin{minipage}{0.5\hsize}
   \begin{center}
    \includegraphics[width=80mm]{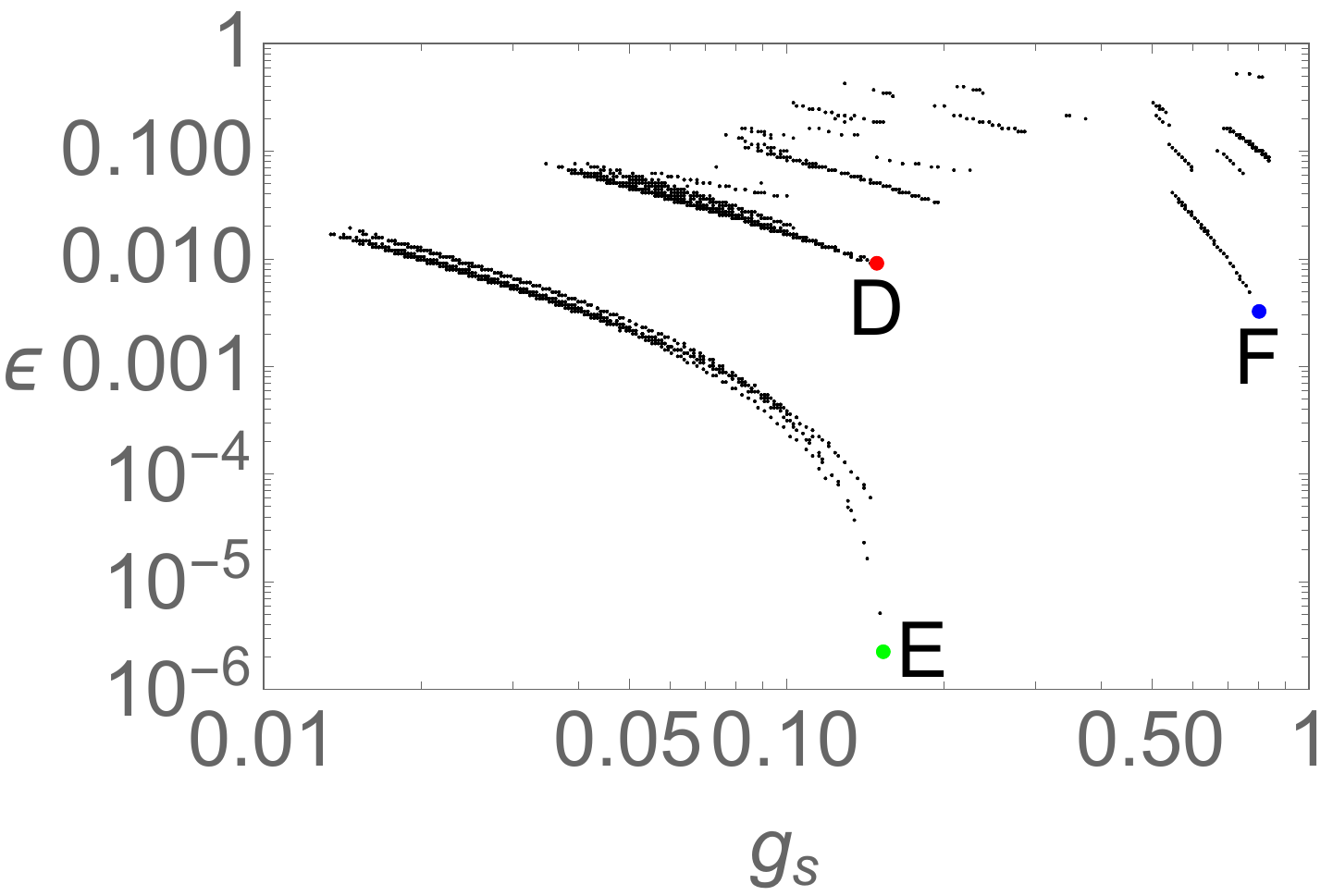}
   \end{center}
  \end{minipage}
    \caption{Numerical results for the distribution of non-supersymmetric F-theory flux vacua.
    Each dot represents a solution with different set of fluxes and the ambiguity from 
    SL(2, $\mathbb{Z}$) duality has been deducted. Note that, as in Figure \ref{fig:T6},  
    these dots can be degenerated in this 2D plot, due to the existence of other directions of the 
    stabilized moduli $z$ and $z_1$. The right figure is the log-log plot of the same results.}
    \label{fig:F}
\end{figure}

\begin{figure}[htb]
  \begin{minipage}{0.5\hsize}
   \begin{center}
     \includegraphics[width=80mm]{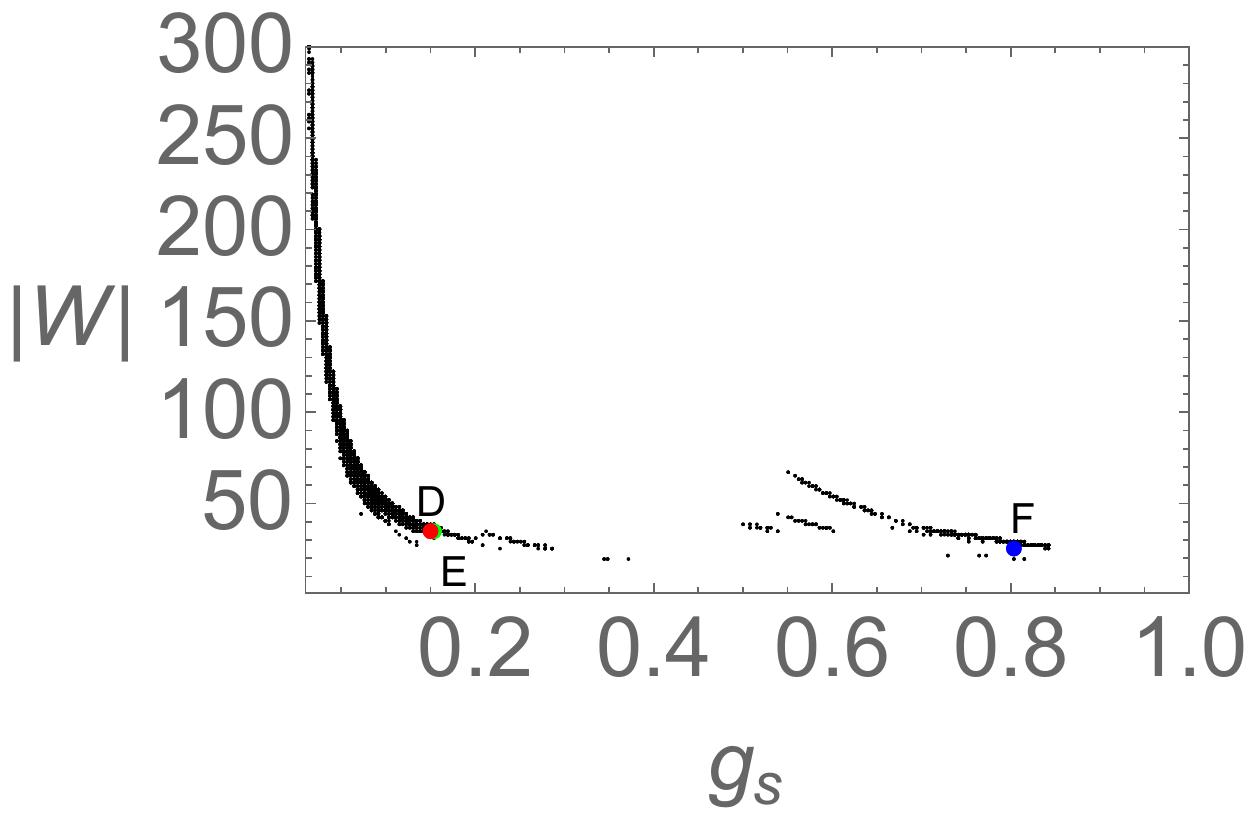}
    \end{center}
   \end{minipage}
  \begin{minipage}{0.5\hsize}
   \begin{center}
    \includegraphics[width=80mm]{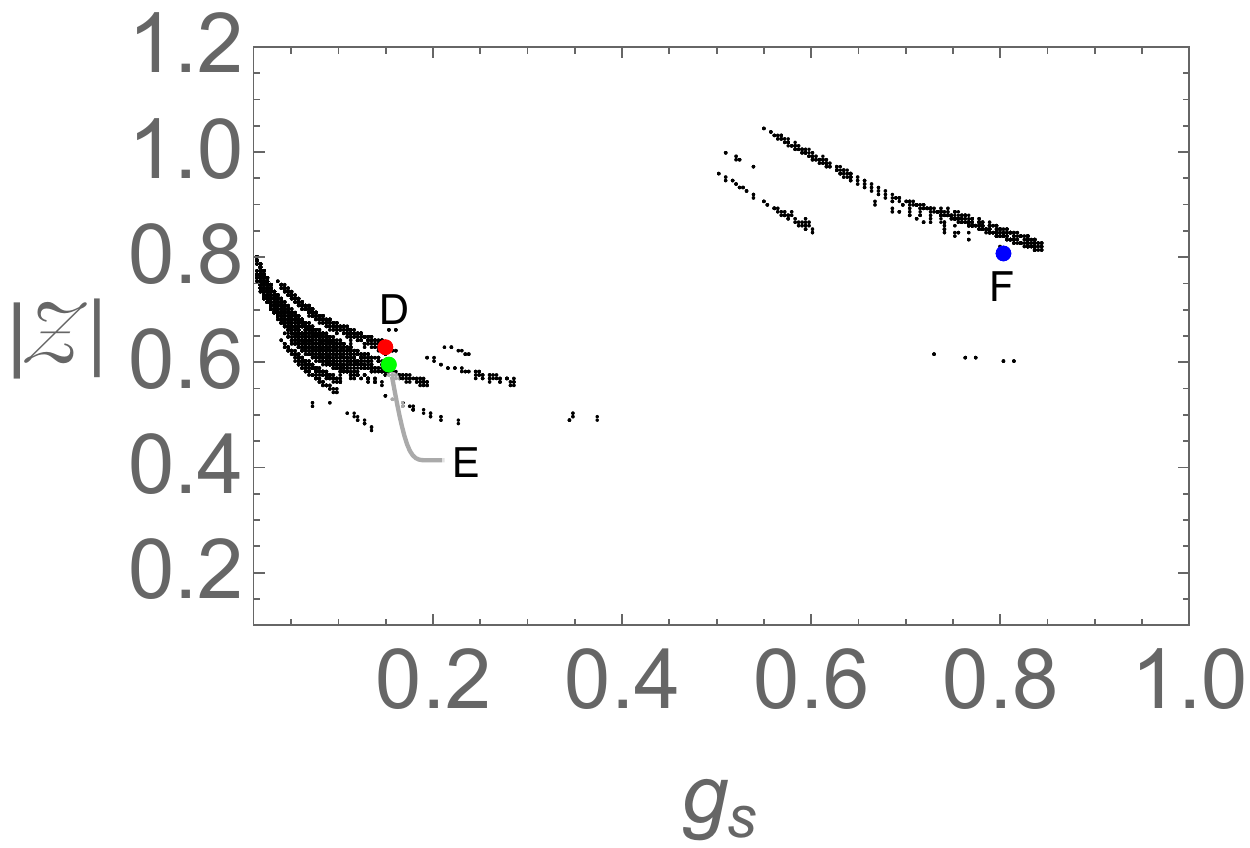}
   \end{center}
  \end{minipage}
    \caption{Numerical results for the absolute values of the flux superpotential $|W|$ and 
    $|{\cal Z}|=e^{K/2}|W|$ on each of the flux vacua depicted in Figure \ref{fig:F}.
    }
    \label{fig:F3}
\end{figure}

As a result, we numerically found 10058 supersymmetry-breaking flux vacua under the following 
range of background fluxes\footnote{By using the Monte-Carlo methods, we randomly generated $10^6$ 
data set of fluxes within $-10\leq n_{1,2,\cdots,11}\leq 10$ and confirmed that the possible minima 
are almost clustered around $n_1=n_2=0$. While this indication makes our present artificial choice 
of fluxes due to the limited performance of our computers quite conceivable, it would be instructive 
to re-analyze the model in a broader range of fluxes.} 
\begin{align}
n_1=n_2=0, \quad -8\leq n_{3,4,8,9,10,11}\leq 8, \quad -10\leq n_{5,6} \leq 10, \quad -6\leq n_7\leq 6,
\label{eq:flux_F}
\end{align}
satisfying the tadpole cancellation condition~(\ref{tadcanF}). Our numerical results are summarized in 
Figures \ref{fig:F} and \ref{fig:F3}. 
Note that we have not drawn the area over a range $0.6 < \epsilon <1$, where we confirmed that there 
are no appropriate solutions in our present setup. 
As advocated in the previous section, one can see clearly that the scalar potential and other 
on-shell quantities thus obtained exhibit a monotonic, non-increasing functional behavior over 
a wide range of parameter $g_s$. One can also check that all the numerical solutions in 
Figure \ref{fig:F} satisfy the condition (\ref{eq:epsilon}), if we take a sufficiently large 
volume ${\cal V}$, e.g. larger than $\simeq {\cal{O}}(10^2)$ for the background geometry.

\clearpage
Here we exemplify our numerical results by picking up three independent flux vacua determined by 
the following set of fluxes: 
\begin{table}[htb]
    \centering
    \begin{tabular}{|c|c|} \hline
   Vacuum    & Set of fluxes $(n_1, n_2 n_3, n_4, n_5, n_6, n_7, n_8, n_9, n_{10}, n_{11})$ \\ \hline \hline
   D   & $(0,0, -2, -8, -8, -2, -6, -4, -7, -8, 0)$ \\ \hline
   E    &  $(0,0, -3, -7, 4, 1, -2, -7, -7, -8, -2)$ \\ \hline
   F    & $(0,0, -4, 1, -10, -10, 2, 7, -8, 6, 0)$ \\  \hline
    \end{tabular}
\end{table}

\noindent These three examples D, E and F are represented in Figure \ref{fig:F} by the red, 
green and blue points, respectively. Explicit values of stabilized moduli and various on-shell 
quantities at the each of the vacua are summarized in Tables~\ref{tab:F1} and~\ref{tab:F2}.\footnote{
More precisely, at the each of the minima, we found that the VEVs of moduli fields exhibit the following distributions:
\begin{align*}
1 < {\rm Im}z_1 < 1.34, \ \ \ \ 2<  {\rm Im}z < 4.54, \ \ \ \ 1.19<  {\rm Im}S < 74.6.
\end{align*}
Naively the smallness of ${\rm Im}z_1$ may suggest that all the numerical solutions we have obtained are likely to be in a region 
outside the ``parametrically large field distance" forbidden by the swampland hypothesis \cite{Obied:2018sgi,Garg:2018reu,Ooguri:2018wrx} 
and our results are consistent with the recent developments on this field. We would like to thank the referee for making us realize this point.}
Comparing these results as well as the plots of Figure \ref{fig:F} in F-theory setup with those 
in Type IIB orientifold model in Section \ref{subsec:resultIIB}, it appears that within a finite 
range of background fluxes of the same amount, the model based on F-theory flux compactification 
equipped with a next-to-leading-order $g_s$ correction prefers a smaller value of the vacuum energy.
This preference would provide strong motivation for further studies toward the de Sitter model 
building in the framework of F-theory dealing with the finite string coupling corrections.

\begin{table}[ht]
    \centering
    \begin{tabular}{|c|c|c|c|c|c|} \hline
   Vacuum    &  $z_1$  & $z$ & $S$ & $g_s$ & $\epsilon$ \\ \hline \hline
   D    & $-0.703+1.00 i$ & $-0.0908 + 2.08 i$ & $-0.275 + 6.74i$ & 0.148 & $9.37\times 10^{-3}$ \\ \hline
   E   & $-0.181+1.01 i$ & $0.0584 + 2.15 i$ & $-0.461 + 6.51i$ & 0.154 & $2.16\times 10^{-6}$ \\ \hline
   F     & $0.243 + 1.01 i$ & $0.435 + 2.03i$ &  $-0.354 + 1.25 i$ & 0.802 & $3.30\times 10^{-3}$ \\ \hline   
    \end{tabular}
    \caption{Explicit values of stabilized moduli and on-sell quantities in $M_{\rm Pl}=1$ unit. 
    }
    \label{tab:F1}
\end{table}
\begin{table}[ht]
    \centering
    \begin{tabular}{|c|c|c|} \hline
   Vacuum    & $W$ & Eigenvalues of mass matrix $\partial_I \partial_J V \times {\cal V}^2$ \\ \hline \hline
   D     & $16.9-32.0i$  &   (5.13, 2.13, 1.85, 1.26, 0.0446, $2.01\times 10^{-4}$)\\ \hline
   E    & $-9.16-33.8i$  &  (3.99, 3.83, 2.89, 1.42, 0.0393, $1.54\times 10^{-6}$ )\\ \hline
   F    & $25.9+2.33i$   &  (7.24, 4.56, 3.27, 0.509, 0.154, $7.06\times 10^{-3}$)\\ \hline
    \end{tabular}
    \caption{Explicit values of flux superpotential and mass eigenvalues in $M_{\rm Pl}=1$ unit. Note that 
    we have diagonalized the mass matrix and the eigenvalues are displayed in the descending order.}
    \label{tab:F2}
\end{table}

Before ending this section, we would like to discuss about a possible extension of our F-theory 
setup toward the stabilization of K$\ddot{{\textrm{a}}}$hler moduli and explicit construction of 
de Sitter spacetime. As can be seen from Figure \ref{fig:F3}, the absolute value of the 
superpotential is of, at lowest, ${\cal O}(10)$. This implies that LVS-type extension is more 
desirable to construct an explicit de Sitter spacetime, rather than utilizing the KKLT-type 
construction which generically requires an extremely small on-shell value of the flux superpotential.
Suppose we employ the LVS-type argument when we extend the K$\ddot{{\textrm{a}}}$hler moduli sector 
in our F-theory setup, and assume that quantum corrections for K$\ddot{{\textrm{a}}}$hler moduli 
fields breaking the no-scale structure give rise to a potential
\begin{align}
V \sim -e^{
-\ln \left( 2\widetilde{Y} {\rm Im}S \right)}|W_0|\frac{\ln {\cal V}}{{\cal V}^3}.
\end{align}
Then the dimensionless constant $\epsilon$ determined in the dynamics of complex structure moduli 
sector is further constrained to satisfy a condition $\epsilon\sim e^{K_{\rm CS}}|W_0|{\cal V}^{-1}
\ln {\cal V}$ such that the total effective potential becomes close to zero as a candidate of the 
tiny cosmological constant. Since the numerical value of $\epsilon$ obtained in our analysis is 
sufficiently small to realize this naive expectation, it would be interesting to cast this idea in 
a rigorous setup.\footnote{For the details about quantum corrections in F-theory, we refer the 
reader to, for instance, \cite{Grimm:2013gma,Grimm:2013bha,Minasian:2015bxa,Weissenbacher:2019mef}.}

\section{Flux Vacua and Attractor Equations}
\label{sec:Att}

So far we have analyzed the vacuum structure of effective theories of moduli fields arising 
from Type IIB and F-theory flux compactifications, aiming to construct appropriate 4D spacetime 
describing our universe in a consistent and systematic way. There we have shown that the on-shell 
quantities such as vacuum energy and flux superpotential exhibit a monotonic, non-increasing 
functional behavior with respect to the string coupling. While this indication strongly motivates 
us to devote special attention to the significance of F-theory framework for the de Sitter model 
building, at the same time the following natural question arises: how exactly and why these 
on-shell quantities on the flux vacua in moduli stabilization problem are determined to exhibit
this fascinating behavior?

Here we argue that our findings are deeply connected with a seemingly different topic in the 4D 
supergravity referred to as the attractor mechanism \cite{Ferrara:1995ih} (see 
also \cite{Ferrara:2008hwa,Moore:2004fg}), where the normalized flux superpotential ${\cal Z}=e^{K/2}W$ 
of a system can be identified as a central charge of the underlying algebra which is subject to a 
monotonic flow along the radial direction of a background. 
More precisely, the attractor mechanism means that in a class of extremal 
black holes in 4D ${\cal N}=2$ theories, moduli fields are drawn to fixed values at the horizon, 
regardless of the initial values at the asymptotic infinity. There the fixed values of the moduli 
fields are determined by the so-called attractor equation. In fact, a possible existence of a 
correspondence between flux vacua in moduli stabilization and the attractor mechanism in 
supergravity has been pointed out in \cite{Kallosh:2005ax} (see 
also \cite{Kallosh:2005bj,Kallosh:2006bt,Alishahiha:2006jd,Bellucci:2007ds,Larsen:2009fw}). From this perspective, 
the systematic behavior of various quantities at the distributed flux vacua we observed may 
reflect the dynamics of a corresponding attracting object, if this kind of correspondence truly exists.
Here we provide a supporting evidence for this conjecture by confirming that various solutions we 
obtained indeed satisfy the suitably generalized attractor equations simultaneously.

First let us reconsider the  $T^6/\mathbb{Z}_2$ toroidal orientifold model as a limit of F-theory on an uplifted fourfold $T^6 \times T^2/\mathbb{Z}_2$, in accordance with a prescription illustrated in \cite{Kallosh:2005ax,Kallosh:2005bj}. 
Seen from this perspective, the 
K$\ddot{{\textrm{a}}}$hler potential and flux superpotential in the isotropic case with moduli 
fields $\Phi_I = \{ \tau, S \}$ in Section \ref{subsec:example1} can be derived from the generic 
formulas (\ref{Kper}) and (\ref{fluq})  in F-theory compactifications with the following assignment 
of the fourfold periods
\begin{align}
\begin{split}
\Pi_1 &=1, \ \Pi_2=-S, \ \Pi_3=\tau, \ \Pi_4=-S\tau, \\
\Pi_5 &=3\tau^2, \ \Pi_6=-3S\tau^2, \ \Pi_7=-\tau^3, \ \Pi_8=S\tau^3,
\end{split}
\label{peritoro}
\end{align}
and the intersection matrix
\begin{align}
\eta=\begin{pmatrix}
0 & 0 & 0 & 0 &  0 & 0 & 0 & 1 \\
0 & 0 & 0 & 0 &  0 & 0 & -1 & 0 \\
0 & 0 & 0 & 0 &  0 & 1 & 0 & 0 \\
0 & 0 & 0 & 0 &  -1 & 0 & 0 & 0 \\
0 & 0 & 0 & -1 &  0 & 0 & 0 & 0 \\
0 & 0 & 1 & 0 &  0 & 0 & 0 & 0 \\
0 & -1 & 0 & 0 &  0 & 0 & 0 & 0 \\
1 & 0 & 0 & 0 &  0 & 0 & 0 & 0
\end{pmatrix},
\label{intmtoro}
\end{align}
where the background three-form fluxes in the $T^6/\mathbb{Z}_2$ model can be reproduced from (\ref{Gflux}) 
with
\begin{align}
\vec{n} = \int G_4=(-c^0, a^0, -c^1, a_1, -3d_1,-3b_1,d_0,-b_0)^T.
\end{align}
The above reformulation corresponds to geometrize the SL(2, $\mathbb{Z}$) duality of Type IIB string 
theory into an auxiliary two-torus and combine the various three-forms in Section \ref{subsec:example1} 
with additional cohomology classes dual to the A- and B-cycles of the torus.\footnote{See for instance 
Section 3 in \cite{Denef:2008wq} for more details and precautions.}

By using above quantities and the central charge ${\cal{Z}}(\tau,\bar{\tau},S,\bar{S}) = e^{\frac{K}{2}}
W(\tau,S)$, one can check that the Minkowski solutions given by (\ref{torvacuum}) and (\ref{torvac2}) 
of the effective potential also satisfy the attractor equation of a generalized 
form \cite{Kallosh:2005ax,Denef:2004ze}
\begin{align}
n_i = 2 {\textrm{Re}} \left[ \overline{{\cal{Z}}}\hat{\Pi}_i + \overline{D}^{\tau} \overline{D}^S
\overline{{\cal{Z}}} D_{\tau} D_S \hat{\Pi}_i \right],
\end{align}
where $\overline{D}^I \equiv K^{I\bar{J}}{\overline{D}}_{\bar{J}}$ and we have 
K$\ddot{{\textrm{a}}}$hler-normalized the periods as $\hat{\Pi}_i \equiv e^{K/2}\Pi_i$. The appearance 
of this equation can be also understood from the viewpoint of the Hodge structure of allowed four-forms 
of the system \cite{Denef:2004ze}. In a similar fashion, we numerically checked that the 
non-supersymmetric flux vacua with nonzero scalar potential described in Section \ref{subsec:resultIIB} 
also satisfy the equation
\begin{align}
n_i = 2 {\textrm{Re}} \left[ \overline{{\cal{Z}}}\hat{\Pi}_i - \overline{D}^I \overline{{\cal{Z}}} 
D_I \hat{\Pi}_i + \overline{D}^{\tau} \overline{D}^S
\overline{{\cal{Z}}} D_{\tau} D_S \hat{\Pi}_i \right].
\end{align}
More precisely, we have picked up several points in Figure \ref{fig:T6} as well as the points A, B and C 
and confirmed that in all the examples the equation indeed holds up to the accuracy ${\cal{O}}(10^{-20})$, 
as far as we investigated.\footnote{It is worth performing further consistency checks about the appearance 
of this equation for non-supersymmetric flux vacua from another independent way, either by analytical or 
more sophisticated numerical approach.}

Before moving on to the analysis about F-theory example, we would like to mention that a whole picture of 
a possible correspondence between flux vacua and attractor mechanism is not completely elucidated even in 
a simple situation in Type IIB compactification as we actualized above. The main problem is the lack of 
knowledge about the identification of the exact metric of a corresponding black object, from which the 
attractor behavior and the entropy can be extracted to confirm the coincidence with the analysis of the 
flux vacua. Although we naively expect that a warped geometry in 10D spacetime with appropriate fluxes and 
horizon topology would be relevant, a detailed study of this subject is beyond the scope of the present 
paper. The simple $T^6/\mathbb{Z}_2$ toroidal orientifold with a suitable set of background fluxes we 
thoroughly investigated here may be regarded as a moderate starting point to address this important problem.

Finally, let us discuss about more nontrivial check concerning the F-theory model discussed in 
Section \ref{sec:Fth}, equipped with the moduli fields $\Phi_I = \{ z, z_1, S \}$. After a careful 
calculation utilizing the fourfold periods (\ref{per}) and the central charge ${\cal{Z}}(z, \bar{z}, z_1, 
\bar{z_1}, S, \bar{S}) = e^{\frac{K}{2}}W(z, z_1, S)$, one can check that the Minkowski solution 
in (\ref{vacuum}) simultaneously satisfies the attractor equation of the following form
\begin{align}
n_i = 2 {\textrm{Re}} \left[ \overline{{\cal{Z}}}\hat{\Pi}_i + C^{IJ} D_I D_J \hat{\Pi}_i \right],
\label{Fatt}
\end{align}
where $C^{IJ}$ is determined by a condition\footnote{See Appendix \ref{app:attF} for the details about 
our F-theory generalization of the attractor equation.}
\begin{align}
\overline{D}_{\bar{K}} \overline{D}_{\bar{L}} \overline{Z} = C^{IJ} \biggl[ R_{I\bar{K}J\bar{L}}
+K_{I\bar{K}}K_{J\bar{L}} +K_{I\bar{L}}K_{J\bar{K}}\biggl] 
+ \overline{C}^{\bar{I}\bar{J}} e^K \overline{Y}_{\bar{I}\bar{J}\bar{K}\bar{L}}.
\label{Ccon}
\end{align}
Here $Y_{IJKL}$ represent the classical quadruple intersection numbers of the background Calabi-Yau fourfold whose explicit values can be easily extracted from (\ref{FY}), and $R_{I\bar{J}K\bar{L}}$ 
is the Riemann curvature tensor of the complex structure moduli space defined by
\begin{align}
R_{I\bar{J}K\bar{L}} = K^{M\bar{N}}(\partial_{\bar{L}}\partial_{\bar{J}}\partial_M K)
\partial_{I}\partial_{\bar{N}}\partial_K K -\partial_{\bar{L}}\partial_I\partial_{\bar{J}}\partial_K K.
\end{align}
Note that the covariant derivative $D_I$ appearing in (\ref{Fatt}) and (\ref{Ccon}) is not a simple 
K$\ddot{{\textrm{a}}}$hler covariant derivative, but also includes the information about the curvature 
of the moduli space as symbolically represented by $D = \partial + a \partial K + \Gamma.$ Here $a$ is 
the K$\ddot{{\textrm{a}}}$hler weight\footnote{For instance, ${\cal{Z}}$ and $\overline{{\cal{Z}}}$ have 
K$\ddot{{\textrm{a}}}$hler weights $1/2$ and $-1/2$, respectively.} of a quantity the covariant derivative 
acts on, and $\Gamma$ denotes the Christoffel symbol of the second kind of the moduli space metric. For 
the case of $T^6/\mathbb{Z}_2$ model discussed above, $\Gamma$ is exactly zero and therefore can be 
identified with the ordinary K$\ddot{{\textrm{a}}}$hler covariant derivative. 

Moreover, we numerically checked that, up to the accuracy ${\cal{O}}(10^{-20})$, the 
non-supersymmetric flux vacua with nonzero scalar potential found in Section \ref{subsec:resultF} satisfy 
the equation
\begin{align}
n_i = 2 {\textrm{Re}} \left[ \overline{{\cal{Z}}}\hat{\Pi}_i - \overline{D}^I \overline{{\cal{Z}}} 
D_I \hat{\Pi}_i+ C^{IJ} D_I D_J \hat{\Pi}_i \right],
\end{align}
where the coefficient $C^{IJ}$ is constrained to satisfy (\ref{Ccon}). These results strongly support a 
possible existence of a F-theory generalization of the attractor mechanism, going beyond the conventional 
Type IIB setup. Especially, all the explicit results we have shown in the F-theory setup should be regarded 
as a prediction that if there exists a black object exhibiting the attracting behavior in the framework 
of F-theory dealing with finite string coupling, its Bekenstein-Hawking entropy and other on-shell quantities 
would have the same characteristics we have demonstrated above.

\section{Conclusions and Discussions}
\label{sec:conclusion}

In this paper, we studied several aspects of moduli stabilization in the framework of spacetime 
flux compactifications by explicitly constructing flux vacua and analyzing their characteristics. 
Especially we thoroughly investigated the 4D effective theories arising from Type IIB string 
theory on a toroidal orientifold $T^6/\mathbb{Z}_2$ and F-theory on a Calabi-Yau fourfold with 
appropriately quantized background fluxes. As well as clarifying the Minkowski minima with vanishing
scalar potential analytically, we numerically solved the extremal conditions of the moduli fields 
and confirmed that the richness of the choice of background fluxes leads to the existence of vast 
numbers of stable non-supersymmetric minima of the potential with sufficiently tiny vacuum energy. 

Moreover, it turned out that the allowed values of the potential minima and various on-shell 
quantities show a non-increasing functional behavior with respect to the string coupling $g_s$. 
In particular it appears that within a finite range of background fluxes of the same amount, 
F-theory model equipped with a next-to-leading-order $g_s$ correction prefers a smaller value of 
the vacuum energy, compared with the simple Type IIB toroidal orientifold model. We hope that 
several extensions of our results would enrich the stringy construction of the de Sitter spacetime. 

We also argued that an interesting conclusion follows from the suggestive behavior of on-shell 
quantities of Type IIB string and F-theory flux vacua. In the process of clarifying the underlying 
dynamics of the non-increasing functional behavior of the vacuum energy and the superpotential, we 
come up with a conclusion that the previously-reported possible correspondence between flux vacua 
in moduli stabilization problem and attractor mechanism in supergravity lies behind. To provide a 
nontrivial evidence for this surmise, we have also checked that our analytic and numerical solutions 
in various flux backgrounds indeed satisfy the suitably generalized attractor equations simultaneously.
Especially, our demonstrations in F-theory framework go beyond the conventional Type IIB  setups 
describing the attractor mechanism and may shed new light on this intriguing subject.

Finally, we comment on several possible future research directions in a random order.
\begin{itemize}
\item
Our studies and demonstrations about the vacuum structure of effective theories would be straightforwardly 
applicable to broader class of examples based on other type of toroidal orientifolds such as 
$T^6/(\mathbb{Z}_2 \times \mathbb{Z}_2)$ \cite{Blumenhagen:2003vr,Cascales:2003zp} and various 
Calabi-Yau manifolds. Besides providing concrete examples of moduli stabilizations toward the explicit 
de Sitter model building, this would be connected with a further development of the attractor mechanism 
in various setups. Within this context, the recent works about asymptotic flux compactifications 
in \cite{Grimm:2019ixq} and attractor black holes in \cite{Hulsey:2019xdb} can be also involved. 

\item
Since our analysis has been restricted to the complex structure moduli as well as the dilaton in Type IIB language, it is indispensable to address the dynamics of the K\"{a}hler moduli and their stabilization at the quantum level if one wants to find a landscape of explicit de Sitter minima. As has been emphasized for instance in \cite{Dasgupta:2014pma}, it will be required to perform a careful analysis with respect to the consistency with the Einstein equations of motion in order to realize the de Sitter spacetime correctly, which should be a highly nontrivial task.

\item
We naively guess that the standard monotonic gradient flow of the central charge of ordinary attracting extremal BH with respect to the radial direction of a background in SUGRA might be extended at the finite string coupling regime such that the central charge or its derivatives related to the vacuum energy will become also a function of string coupling, whose dependence becomes monotonic if the radial evolution or energy scale in AdS/CFT context and the string coupling are also monotonically correlated. 
Clarifying this point would be an interesting but highly difficult problem.

\item
The attractor mechanism states that the evolution of moduli fields is governed by the gradient flow in 
the radial direction of the 4D background and the central charge ${\cal{Z}}$. On the other hand, it has 
been recently pointed out that there exists a close relation between the swampland argument and the 
gradient flow in gravity \cite{Kehagias:2019akr}. Therefore, naively one can expect that our findings 
about the appearance of the attractor equations in both of the Type IIB string and F-theory flux 
compactifications might be connected with the basics of the swampland hypothesis. It would be interesting 
to figure out a more complete description about this expectation.

\item
About KKLT construction, recent analysis in \cite{Bena:2018fqc} about the conifold modulus parametrizing 
the size of three-cycle on the bottom of a warped deformed conifold throat indicates that one must 
require a large amount of three-form background fluxes to avoid the destabilization, which is generically 
incompatible with the tadpole cancellation condition in Type IIB flux compactifications. It would be 
interesting to clarify whether F-theory framework we studied can provide one way to resolve this 
destabilization problem. 

\end{itemize}

\subsection*{Acknowledgements}

We would like to thank K. Ohta, A. Otsuka, K. Sakai and T. Watari for useful discussions and 
comments. H. O. was supported by a Grant-in-Aid for JSPS Research Fellow No. 19J00664. We 
appreciate the Yukawa Institute for Theoretical Physics at Kyoto University, where this work 
was presented during the YITP-W-19-10 on ``Strings and Fields 2019". We are also grateful to 
the participants of the workshop ``KEK Theory Workshop 2018" held at KEK Theory Center for 
illuminating discussions.
\appendix

\section{Hodge Structure and Attractor Equation}\label{app:attF}

We have described a F-theory generalization of the attractor equation in (\ref{Fatt}) with a 
condition (\ref{Ccon}). Here let us explain how we arrived at these expressions. In fact, a 
slight modification of the analysis in \cite{Denef:2004ze} leads us to find out appropriate 
form of the equation.

Along the line of \cite{Denef:2004ze} (see also \cite{Bellucci:2007ds}), here we consider an 
expansion of a generic real four-form flux on a Calabi-Yau fourfold $X_4$ as\footnote{Although 
the third and forth order covariant derivatives of $\hat{\Omega}$ are irrelevant to derive 
(\ref{Fatt}), it would be instructive to clarify their relationship to other basis and obtain 
a more complete view of the Hodge structure of the system.}
\begin{align}
G_4 &= A \hat{\Omega} + B^I D_I \hat{\Omega} +C^{IJ} D_I D_J \hat{\Omega}+\overline{A} 
\overline{\hat{\Omega}} +{\overline{B}}^{\bar{I}} \overline{D}_{\bar{I}}\overline{\hat{\Omega}} 
+{\overline{C}}^{\bar{I}\bar{J}} \overline{D}_{\bar{I}}\overline{D}_{\bar{J}}\overline{\hat{\Omega}}, 
\label{eq:G4expand}
\end{align}
where the indices $I, J$ label the complex structure moduli of a Calabi-Yau fourfold $X_4$, 
and we have K$\ddot{{\textrm{a}}}$hler-normalized the holomorphic four-form as $\hat{\Omega} 
(\Phi_I,\overline{\Phi}_{\bar{I}}) \equiv e^{K/2}\Omega(\Phi_I)$ such that the covariant derivative 
would acts on $\hat{\Omega}$ as $D_I\hat{\Omega}=(\partial_I +K_I/2)\hat{\Omega}=e^{K/2}D_I\Omega$. 
Let us denote the intersection products of the covariant derivatives of $\hat{\Omega}$ as
\begin{align}
\begin{split}
{\cal N}_{I...J|K...L} &\equiv D_I \cdots D_J \hat{\Pi}\; \eta^{-1} D_K\cdots D_L \overline{\hat{\Pi}} 
= \int_{X_4} D_I\cdots D_J \hat{\Omega} \wedge D_K \cdots D_L \overline{\hat{\Omega}}, \\
{\cal N}^\prime_{I...J|K...L} &\equiv D_I \cdots D_J \hat{\Pi}\; \eta^{-1} D_K\cdots D_L \hat{\Pi} 
= \int_{X_4} D_I\cdots D_J \hat{\Omega} \wedge D_K \cdots D_L \hat{\Omega},
\end{split}
\end{align}
where we have K$\ddot{{\textrm{a}}}$hler-normalized the fourfold periods of $X_4$ as $\hat{\Pi} 
\equiv e^{K/2}\Pi$. Their explicit forms are known to be given by (see for 
instance \cite{Denef:2004ze})
\begin{align}
\begin{split}
{\cal N}_{I|\bar{J}} &= -K_{I\bar{J}}, \\
{\cal N}_{IJ|\bar{K}\bar{L}} & = R_{I\bar{K}J\bar{L}} +K_{I\bar{K}}K_{J\bar{L}} +K_{I\bar{L}}K_{J\bar{K}}, \\
{\cal N}_{\bar{I}J|K\bar{L}} &= K_{J\bar{I}}K_{K\bar{L}}, \\
{\cal N}^\prime_{IJ|KL} &= e^K Y_{IJKL},
\end{split}
\end{align}
where $Y_{IJKL}$ represent classical quadruple intersection numbers of $X_4$, and $R_{I\bar{J}K\bar{L}}$ 
is the Riemann curvature tensor of the complex structure moduli space of $X_4$.

Utilizing the above formulas, it turns out that the explicit forms of the coefficients $A$ and $B^I$ 
in (\ref{eq:G4expand}) can be determined by
\begin{align}
\begin{split}
\overline{{\cal{Z}}} &= \int G_4 \wedge \overline{\hat{\Omega}} =  A \int \hat{\Omega} \wedge 
\overline{\hat{\Omega}} = A, \\
\overline{D}_{\bar{I}}\overline{{\cal{Z}}} &= \int G_4 \wedge \overline{D}_{\bar{I}} \overline{\hat{\Omega}}     
= B^J \int D_J \hat{\Omega} \wedge \overline{D}_{\bar{I}} \overline{\hat{\Omega}} = B^J (-K_{J\bar{I}}), 
\label{ABco}
\end{split}
\end{align}
whereas $C^{IJ}$ and its conjugate are constrained to satisfy
\begin{align}
\begin{split}
\overline{D}_{\bar{K}} \overline{D}_{\bar{L}} \overline{{\cal{Z}}} &= \int G_4 \wedge \overline{D}_{\bar{K}} 
\overline{D}_{\bar{L}} \overline{\hat{\Omega}} \\
&=C^{IJ} \int D_I D_J \hat{\Omega} \wedge \overline{D}_{\bar{K}} \overline{D}_{\bar{L}} \overline{\hat{\Omega}}
+\overline{C}^{\bar{I}\bar{J}} \int\ \overline{D}_{\bar{I}} \overline{D}_{\bar{J}} \overline{\hat{\Omega}} 
\wedge \overline{D}_{\bar{K}} \overline{D}_{\bar{L}} \overline{\hat{\Omega}} \\
&= C^{IJ} \biggl[ R_{I\bar{K}J\bar{L}}+K_{I\bar{K}}K_{J\bar{L}} +K_{I\bar{L}}K_{J\bar{K}}\biggl] 
+\overline{C}^{\bar{I}\bar{J}} e^K \overline{Y}_{\bar{I}\bar{J}\bar{K}\bar{L}}.
\label{CconApp}
\end{split}
\end{align}
By plugging (\ref{ABco}) into (\ref{eq:G4expand}) and regarding (\ref{CconApp}) as an additional condition,
one can obtain the formulas about the attractor equations discussed in Section \ref{sec:Att}. 

In a generic Type IIB Calabi-Yau compactification, the special K$\ddot{{\textrm{a}}}$hler geometry relation 
of an underlying Calabi-Yau threefold yields a simplification of the form $D_iD_j \hat{\Omega} = i Y_{ijk}
K^{k\bar{l}}D_{\bar{l}}\overline{\hat{\Omega}}$ with $Y_{ijk}$ being classical triple intersection numbers 
of the threefold with moduli fields $\Phi_i$. This makes the basis $D_i D_j \hat{\Omega}$ in an analogous 
expansion of a real three-form flux in Type IIB compactifications redundant and as a result the attractor 
equation in conventional Type IIB setups takes a quite simple form. By contrast, the F-theory compactification 
based on a generic Calabi-Yau fourfold does not have such a simple structure of special 
K$\ddot{{\textrm{a}}}$hler geometry and the basis $D_I D_J \hat{\Omega}$ becomes independent. This is the 
main reason why a slight modification to the attractor equation is necessary in F-theory framework.


\end{document}